\def\kms{km~s$^{-1}$}
\def\etal{{\it et al.}}

\def\msun{$M_\odot$}

\documentclass[
    ,final            
  ]
  {aipproc}

\layoutstyle{6x9}

\newcommand\doingARLO[2][]{%
  \ifx\mmref\undefined #1\else #2\fi
}

\begin{document}

\title 
      {Extragalactic HI Surveys at Arecibo: the Future}


\author{Riccardo Giovanelli}{
  address={Space Sciences Bldg., Cornell University, Ithaca, NY 14853},
  email={riccardo@astro.cornell.edu}
}

\copyrightyear  {2008}

\begin{abstract}
Starting in the 1970s, the Arecibo 305m telescope has made seminal contributions
in the field of extragalactic spectroscopy. With the Gregorian upgrade completed
in the late  1990s, the telescope acquired a field of view. Population of that
field of view with a seven--feed array at L--band (ALFA) increased by nearly one
order of magnitude its survey speed. As a result, much of the extragalactic
astronomy time of the telescope is now allocated to survey projects, 
which are briefly discussed. The next technical development stage for the 305m
telescope is foreseen as that of a 40 beam system that would take advantage
of phased array technology: {\bf AO40}. This would further speed up the
survey performance of the telescope. It is shown how the figure of merit for
survey speed of AO40 would be comparable with that of SKA--precursor
facilities, planned for operation in the next decade. A number of 
scientifically desirable new surveys that would become possible with AO40
are briefly discussed.
\end{abstract}

\date{\today}

\maketitle

\section{Extragalactic Spectroscopy at Arecibo: a Timeline.}
The 21 cm line of neutral Hydrogen was first detected in 1953 in a galaxy other 
than our own. The Arecibo Observatory entered operations a decade later, but it
took another decade and the replacement of the 305m telescope primary mirror
with a finer surface, for 
the field of extragalactic spectroscopy to become established at that facility. 
In the following years, the Arecibo telecope became a protagonist in the field of
extragalactic spectroscopy, with fundamental contributions to the understanding
of galactic structure, the predominance of dark matter in galaxies and the case
of nature vs. nurture in their evolution; the large scale structure of the Universe,
its basic parameters and peculiar velocity field. In the 1990s, a second upgrade
of the 305m telescope corrected for the spherical aberration of its primary,
delivering an extended field of view at its focus. The exploitation of that field of view
by populating it with multifeed arrays opened a new era at Arecibo: that of surveys
which would cover large fractions of the 13,000 square degrees of the accessible sky.
Currently about to start or already underway, extragalactic spectroscopy surveys that 
use the ALFA feed array will deliver results of unprecedented scope and impact in the 
field. A progressively larger fraction of telescope time is allocated to survey work.

\section{Currently Ongoing HI Surveys}

ALFA is a seven-feed focal plane array; surveys enabled by this device started taking 
data in 2005. Among them: AGES (the {\it Arecibo Galactic Environment Survey}) aims at
the investigation of the HI contents in the periphery of nearby galaxies, groups and
clusters; ZOA (the {\it Zone of Avoidance} survey) will identify galaxies near the
Galactic plane, helping overcome the limitations imposed by dust extinction; AUDS
(the {\it Arecibo Ultra Deep Survey}) will carry out a high sensitivity pencil beam
exploration to $z\sim 0.25$. The most ambitious among the HI extragalactic surveys
is ALFALFA (the {\it Arecibo Legacy Fast ALFA} survey)\cite{a1}, which in blind mode will cover
7000 square degrees of sky and will detect more than 25,000 HI sources. While ZOA and 
AUDS will initiate in 2008, ALFALFA and AGES are well under way.

The median $cz$ of the ALFALFA survey is near 8000 \kms, the typical scalelength
of baryonic acoustic oscillations. ALFALFA is the only large--scale HI survey
that samples a fair volume of the Universe (HIPASS' median $cz$ is less than
3000 \kms). Moreover, a large fraction of HIPASS sources suffer from
confusion, making the identification of optical counterparts difficult and
often impossible without follow--up higher resolution HI observations. The
smaller Arecibo beam largely obviates the problem: more than 95\% of ALFALFA
sources can be unambiguously associated with the correct optical counterpart.

\begin{figure}[ht]
\resizebox{1.1\textwidth}{!}{\includegraphics{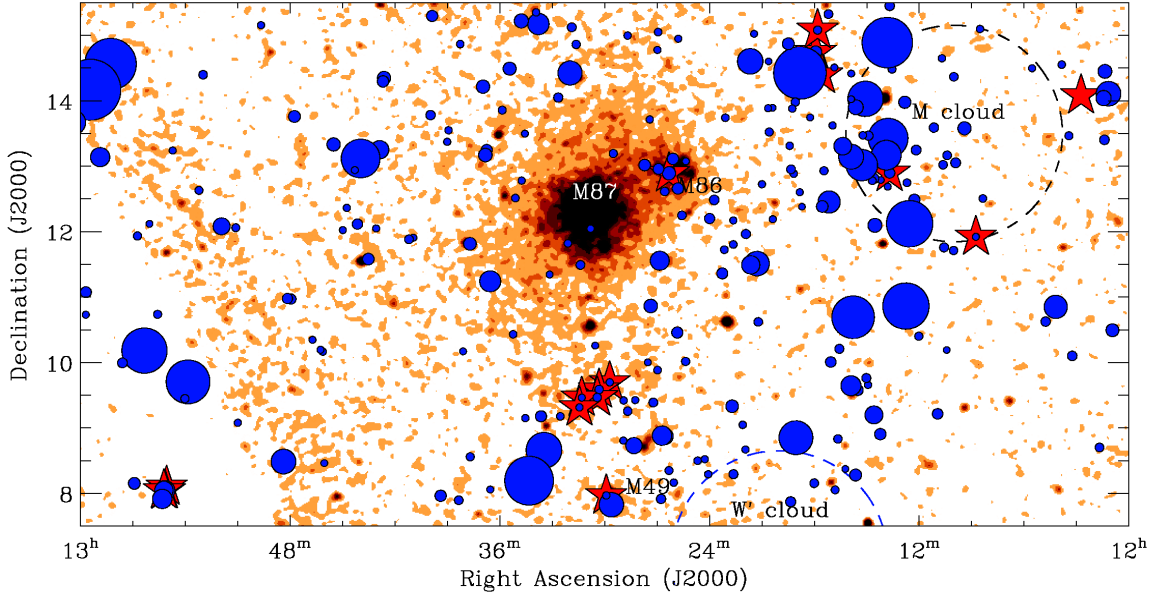}}
\caption{
Composite of the Virgo cluster: X ray emission (orange), HI sources
(blue circles) and HI streams (red stars).
See text for details. 
}
\label{virgo}
\source{Brian Kent}
\end{figure}

The collective science goals of extragalactic HI surveys at Arecibo is 
described in the introductory paper by Giovanelli, of these proceedings,
and several initial findings of those surveys have been
presented. Here, it is worth refreshing our memory on one of such
findings, briefly discussed in the presentation b Brian Kent\cite{kent}.
ALFALFA reports the discovery of huge gas streams in the outer regions 
of the Virgo cluster, extending hundreds of kpc and witnessing to
dramatic events in therevolution of galaxies in those environments.
The northern part of the cluster, now fully mapped by ALFALFA to a
detection limit of $2\times 10^7$ \msun, is shown in Figure \ref{virgo}.
The orange shading shows the extent of the X--ray emission centered on
M87, by hot intracluster gas (ICM); the blue dots are HI detections, their
areas proportional to their HI masses. The picture clearly illustrates the 
impact of the ICM on gas content of galaxies. The red stars show the location
of HI streams, some of which can be tracked over degrees ($1^\circ\simeq 300$ 
kpc at the cluster distance. These features are detected near the column
density limit of ALFALFA ($\sim 5\times 10^{18}$ cm$^{-2}$), suggesting
that ALFALFA --- while allowing us to seeing these structures for the 
first time --- may be seeing just ``the tip of the iceberg''.

\section{AO40}
The Arecibo telescope has the largest collecting area in world radio 
astronomy, about 1/10 of the future SKA and comparable with or larger 
than the SKA precursor telescopes currently under design or construction. 
ALFA has provided a nearly one order of magnitude survey speed, with respect 
to single beam operation. However, the focal plane of the telescope is 
undersampled by about a factor of 16 by ALFA (beam centers are separated by 
approximately 2 beamwidths). An improvement on this state of affairs can
be achieved with the development of focal plane phased arrays (FPPA).
Two important advantages of FPPAs are relevant: (i) they can exploit a larger 
physical area of the focal plane and (ii) beams can be packed together more 
closely than in feed arrays, such as ALFA. A FPPA approaching Nyquist sampling 
of the FOV may allow for 40 instantaneous beams at L--band. We shall refer to 
such an implementation on the Arecibo telescope as ``AO40''. FPPA technology 
is not well developed yet, and it is still uncertain that such devices will be 
able to deliver the low temperaure and broad bandwidth performances we have 
become used to. We underscore that most SKA precursor future telescopes,
such as ASKAP and APERTIF, are betting on the success in implementing such
technology. 

\begin{table}[!t]
\begin{tabular}{lccccccc}
\hline
  & \tablehead{1}{r}{i}{D(m)}
  & \tablehead{1}{r}{b}{Beam(')}
  & \tablehead{1}{r}{b}{$N_{tel}\times N_{beam}$}
  & \tablehead{1}{r}{b}{FoV} 
  & \tablehead{1}{r}{b}{$T_{sys}(K)$}
  & \tablehead{1}{r}{b}{BW(MHz)} 
  & \tablehead{1}{r}{b}{FoM} 
  \\
\hline
AO1     & 225 & 3.5  & $1\times 1$   &  1  & 25 & 100 & 1  \\
ALFA    & 225 & 3.5  & $1\times 7$   &  7  & 30 & 300 & 14 \\
AO40    & 225 & 3.5  & $1\times 40$  & 40  & 50 & 300 & 30 \\
APERTIF & 25  & 30   & $14\times 25$ & 1800& 50 & 300 & 33 \\
ASKAP   & 12  & 60   & $30\times 30$ & 8800& 50 & 300 & 37 \\

\hline
\end{tabular}
\caption{Parameters for the Computation of Survey Speed}
\label{tab1}
\end{table}

Using a standard expression for the survey speed figure of merit, $FoM$, of
a telescope of effective aperture $A_{eff}$, system temperature $T_{sys}$,
total instantaneous solid angle of the survey $\Omega_{fov}$ and bandwidth 
$BW$:
\begin{equation}
FoM\propto (A_{eff}/T_{sys})^2 \Omega_{fov} BW,
\end{equation}
and using the same expectation values for $T_{sys}=50$ K and $BW=300$ MHz
for FPPAs at AO40, ASKAP and APERTIF, the relative $FoM$ for those 3
systems with respect to the single beam Arecibo are respectively 30, 33 
and 37, as shown in Table \ref{tab1}. The $FoM$ for ALFALFA is 5.
 {\it With the implementation of FPPA
technology on which future facilities rely on, Arecibo (AO40) can be as
fast a survey machine as any currently planned}. 
The main disadvantages of Arecibo with respect to those precursor experiments
are that it currently has a limited field of view and that it will always 
have worse angular resolution than a comparable collecting area with a distributed 
aperture. The paramount advantage is, of course, that Arecibo already exists
and is a well understood, efficient photon bucket.

Is there a niche for Arecibo as a competitive HI survey  machine?
The density of galaxies per unit solid angle increases obviously with $z$.
At $z\sim 0.1$, confusion within the $\sim 4'$ HPFW of the Arecibo beam cannot 
often be resolved, unless high quality optical redshifts are available for
the sources within the beam, so that confusion can be overcome kinematically.
This limits the pursuit of blind HI surveys with the Arecibo telescope at high 
$z$. A practical rule of thumb indicates that the median centroiding accuracy 
of a source smaller than the telescope beam, detected
at signal--to--noise $StN$ with a beam of HPFW $\beta$, is approximately
$\beta/StN$. A source detected at a marginal $Stn\simeq 6$ will, at 21 cm
wavelength, have an associated positional error somewhat larger than 0.5'.
Identification of the optical counterpart of an HI source without the benefit 
of an accurate optical redshift at even modest $z~\sim 0.1$ thus becomes 
doubtful. The maximum distance at which a given HI mass can be detected 
increases only as $t_s^{1/4}$, where $t_s$ is the dwell time per beam.
The highest $z$ of any source detected in HI emission is $0.28$, achieved by
Catinella and collaborators at Arecibo\cite{cat}. Detection of even 
the brightest galaxies at such 
redshift requires dwell times of several hours at Arecibo, and much longer
at any other telescope. The survey niche for AO40 is thus that of $z\sim 0$,
deep surveys. A number of those have especially strong astrophysical appeal.

\section{Future Extragalactic Blind HI Surveys.}

A necesary observational activity that will require significant telescope time
after completion of currently ongoing surveys consists in the follow--up
corroboration of interesting but weak sources and deeper mapping of a subset
of such objects. Statistical exercises (e.g. \cite{a1})
indicate that $\sim$20\% of the total survey time will be required for
those functions after survey completion. At the current rate of telescope
time allocation, ALFALFA will require 7 calendar years to complete.
The next stage of exploitation of the capabilities of an AO40 facility
may be the extension of such surveys as ALFALFA and AGES to fuller sky 
coverage. The extension of \fbox{ALFALFA++} to cover the 13000 square degrees
of sky accessible with the telescope would require approximately 500
hours of AO40 time. An extension of \fbox{AGES++} to a more extensive set of 
targets, including clusters and groups of galaxies with  $cz<10000$ \kms, 
to cover 3000 square degrees with 300 seconds of dwell time per beam, 
would require 1500 hours with AO40. These would constitute a veritable 
legacy of complete extension.

More interesting, however, would be deeper surveys that would reach
significantly fainter levels of the HI mass function. As mentioned
before, a significant new discovery of ALFALFA consists in the detection
of numerous sources with no optical counterpart and giant streams of HI, 
very near the detection limits of the survey. A survey of the cluster with 
10 times the sensitivity of ALFALFA would be certain to reveal new marvels 
and, bringing the HI detection limit down to $2\times 10^6$ M$_\odot$, would 
provide an unprecedented investigation of the low mass end of the HI mass and 
luminosity functions. A \fbox{Virgo Deep} survey that would cover 200 sq. deg. with 
an effective dwell time of 4000 sec per beam would require 2000 hours with
AO40. Such Virgo Deep survey would deliver genuinely new science and 
provide a fundamental scientific reference.

The $\lambda$CDM paradigm of galaxy formation predicts the presence of 
low mass systems in mini--filaments within cosmic voids. Such structures
have never been detected. The nearest large--scale void in the local
universe lies in the foreground of the Pisces--Perseus supercluster.
With a diameter of order of 60 Mpc, its center is near $cz\simeq 2500$ \kms.
A \fbox {Deep Void} survey that would cover 400 square degrees of the void solid angle
with 10 times the sensitivity of ALFALFA would require some 4000 hours
of AO40 time. Projected along the line of sight to the central regions
of the Local Group (LG), such a survey would permit detection of LG sources
with  as little as $5\times 10^3$ \msun ~and unprecedented column
density sensitivity (near $3\times 10^{17}$ cm$^{-2}$) in resolved
sources in the LG and High Velocity Cloud complexes.

\section{Future Extragalactic Targeted HI Surveys}

HI can be currently detected in emission from extragalactic objects
out to a $z\sim 1/3$, with dwell times of order of 10 ksec at AO.
HI sources at distances $> 200$ Mpc are most unlikely starless, as
the masses required for their detection with AO correspond squarely in
the category of giant galaxies, for which no theoretical expectation
exists of any inhibition to profuse star forming activity. Thus the study
of such objects in the HI line is best done via targeted, rather than 
blind searches.

GASS is an approved, targeted Arecibo survey to obtain complete ``ID 
cards'' for a sample of galaxies in the local Universe ($z<0.04$). It 
uses the SLOAN and GALEX surveys as target finders for objects for which 
cold gas content and kinematics are to be obtained. The collective 
information will be used to measure galaxy parameters, such as masses, 
current and past star formation activity, etc. The survey sample includes 
some 1000 objects, of which a fraction of $\sim$15\% will be provided
by ALFALFA. A possible extension to \fbox{GASS++}, which would include objects
to $z\simeq 0.3$ (a lookback time of about 4 Gyr), would be able to
survey significant predicted changes in scaling laws of disks, 
relying on the uniquely high quality of the kinematic information
in the HI data.

From observations of damped Lyman Alpha systems we know that
$\Omega_{HI}\simeq 10^{-3}$ at $z> 2$; estimated values at $z=0$
are about a factor of 3 to one order of magnitude lower. Theoretical
estimates of the $\Omega_{HI}(z)$ differ substantially (\cite{baugh},\cite{cen},\cite{nagamine}Baugh \etal
~2004; Cen \etal ~(2003); Nagamine \etal ~2005; Millenium run). The
steeper part of the change in $\Omega_{HI}(z)$ may have taken place
between look-back times of 1--7 Gyr. The observational characterization
of that function is of special importance, with respect to SKA
plans for the ``Billion HI Galaxy'' surveys, the measurement of baryon
acoustic oscillations and the elucidation of the nature of dark energy.
As pointed out before, in spite of its large collecting area, sensitivity 
issues and confusion prevent the effective use of the Arecibo telescope 
for observations of individual galaxies at even higher redshift. However,
the Arecibo beam is well matched to the angular size of intermediate
redshift clusters of galaxies. For example, A1835, at $z=0.25$, has
an SZ signature 4' in diameter. Pointed observations would aim at
the detection of the integrated HI emission of the whole cluster.
An HI mass of $10^{12}$ M$_\odot$ spread over a width of 1500 \kms
~could in principle be detectable in less than 10 hours with the
Arecibo telescope. A number of possible complications could make such
an experiment difficult, however: the ability to detect a broad spectral
signature against system standing waves, the increase in the continuum
noise especially by radio galaxies in the cluster itself, easily come
to mind. Yet, the concept offers substantial promise, which may lead
to the characterization of $\Omega_{HI}(z)$ through much of the
interval of cosmic history to which ground--based DLA work is blind ($z<1.6$).
This is an experiment that will require much tender loving care
in the understanding of telescope systematics. A dual beam option
and sophisticated rfi identification/excision techniques would
be required for such a HI Cluster Integral,\fbox{HIClint}, survey.

\begin{theacknowledgments}
This work has been supported by NSF grants AST--0307661,
AST--0435697, AST--0607007. The Arecibo 
Observatory is part of the National Astronomy and Ionosphere Center
which is operated by Cornell University under
a cooperative agreement with the National Science Foundation.
\end{theacknowledgments}

\doingARLO[\bibliographystyle{aipproc}]
          {\ifthenelse{\equal{\AIPcitestyleselect}{num}}
             {\bibliographystyle{arlonum}}
             {\bibliographystyle{arlobib}}
          }
\bibliography{sample}

\begin{thebibliography}{0}

\bibitem{a1}
  Giovanelli \etal ~2005,
  \emph{AJ} 130, 2589
\bibitem{kent}
  Kent, B.R. 2008,
  \emph{these proceedings}
\bibitem{cat}
  Catinella, B. 2008,
  \emph{these proceedings}
\bibitem{baugh}
  Baugh, C. \etal ~2004,
  \emph{NewAR} 48, 1239
\bibitem{cen}
  Cen, R. \etal ~2003,
  \emph{ApJ} 598, 741
\bibitem{nagamine}
  Nagamine, K. \etal ~2005,
  \emph{IAU Symp} 216, 266
\end{thebibliography}

\end{document}